\documentclass[titlepage,twocolumn,superscriptaddress,prl]{revtex4-1}
\usepackage{amsmath}
\usepackage{amssymb}
\usepackage{graphicx}
\usepackage{color}
\usepackage{floatrow}

\newcommand{\RH}{R_\mathrm{MH}}

\newcommand{\rp}{a}

\begin{document}

\title{Maxwell-Hall access resistance in graphene nanopores}

\author{Subin Sahu}
\affiliation{Center for Nanoscale Science and Technology, National Institute of Standards and Technology, Gaithersburg, MD 20899}
\affiliation{Maryland NanoCenter, University of Maryland, College Park, MD 20742}
\affiliation{Department of Physics, Oregon State University, Corvallis, OR 97331}

\author{Michael Zwolak}
\email[\textbf{Corresponding Author:} ]{mpz@nist.gov}
\affiliation{Center for Nanoscale Science and Technology, National Institute of Standards and Technology, Gaithersburg, MD 20899}

\begin{abstract}
The resistance due to the convergence from bulk to a constriction, for example, a nanopore, is a mainstay of transport phenomena. In classical electrical conduction, Maxwell, and later Hall for ionic conduction, predicted this {\em access} or {\em convergence resistance} to be independent of the bulk dimensions and inversely dependent on the pore radius, $a$, for a perfectly circular pore. More generally, though, this resistance is {\em contextual}, it depends on the presence of functional groups/charges and fluctuations, as well as the (effective) constriction geometry/dimensions. Addressing the context generically requires all-atom simulations, but this demands enormous resources due to the algebraically decaying nature of convergence. We develop a finite-size scaling analysis, reminiscent of the treatment of critical phenomena, that makes the convergence resistance accessible in such simulations. This analysis suggests that there is a ``golden aspect ratio'' for the simulation cell that yields the infinite system result with a finite system. We employ this approach to resolve the experimental and theoretical discrepancies in the radius-dependence of graphene nanopore resistance.
\end{abstract}

\maketitle

Ion  transport through pores and channels plays an important role in physiological functions \cite{hille2001,bagal2012,Rasband2010}  and in nanotechnology, with applications such as DNA sequencing \cite{Kasianowicz1996-1,clarke2009,sathe2011}, imaging living cells \cite{hansma1989,korchev1997,panday2015}, filtration \cite{karan2015}, and desalination  \cite{lee2011}, among others. These pores localize the flow of ions and molecules across a membrane, where sensors, for example, nanoscale electrodes for DNA sequencing \cite{Zwolak08,Zwolak05,Lagerqvist06-1,Lagerqvist07-2,Krems09-1,Tsutsui10-1,Chang10-1} , can interrogate the flowing species as they pass through and where functional elements can selectivity regulate the movement of different species (for example, ion types).

In particular, from DNA sequencing \cite{garaj2010,merchant2010,Schneider2010,heerema2016} to filtration \cite{Joshi2014,abraham2017,OHern2014,Jain2015,Surwade2015}, graphene nanopores and porous membranes are one of the most promising materials for applications. Novel fabrication strategies and designs are under development to create large-scale, controllable porous membranes \cite{OHern2014,Jain2015,rollings2016} and graphene laminate devices \cite{Joshi2014,abraham2017}. Moreover, their single atom thickness makes these systems ideal for interrogating ion dehydration \cite{Sahu2017NanoLett, Sahu2017Nanoscale}, which both sheds light on recent experiments on ion selectivity in porous graphene \cite{OHern2014, Jain2015, rollings2016} and will help analyze the behavior of biological pores \cite{Sahu2017NanoLett, Sahu2017Nanoscale}. Dehydration has been predicted to give rise to ion selectivity and quantized conductance in long, narrow pores \cite{Zwolak09, Zwolak10,song2009,richards2012,richards2012quantifying} but the energy barriers are typically so large that the currents are minuscule, which is rectified by the use of membranes with single-atom thickness \cite{Sahu2017NanoLett, Sahu2017Nanoscale}.

\begin{figure}[t]
\includegraphics[width=\columnwidth]{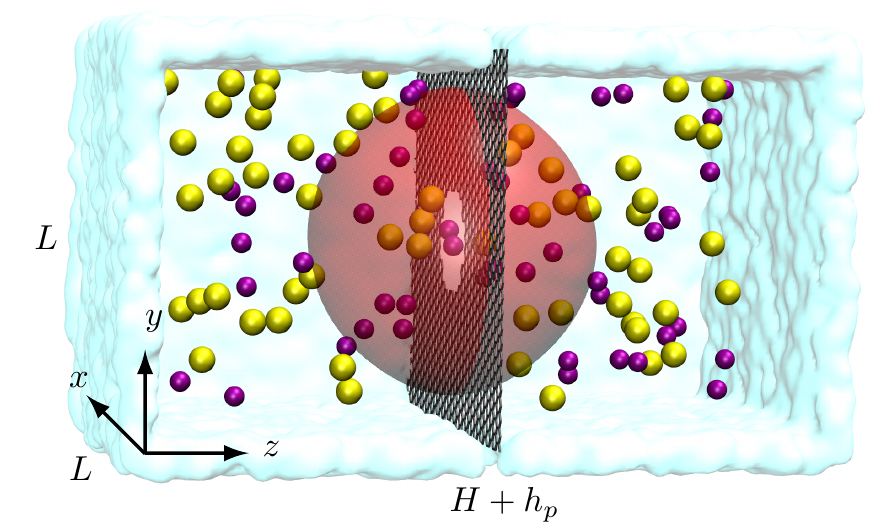}
\caption{\label{Fig1} {\bf Schematic of a graphene nanopore.} The ionic solution is partitioned by a graphene monolayer (the gray, honey-comb membrane) of thickness $h_p$. Potassium (purple) and chloride (yellow) ions are shown as van der Waals spheres but water is not shown even though it is explicitly present in the simulations. The remaining details are in the supplemental material (SM). The red indicates the access region. The total simulation cell is of height $H+h_p$ and cross section $L \times L$.}
\end{figure}

\begin{figure*}[t]
\centering
\includegraphics{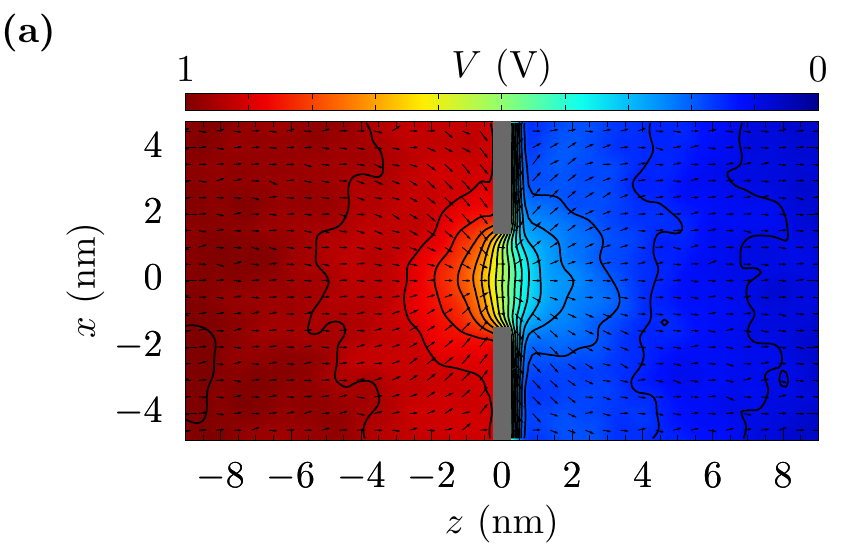}\qquad
\includegraphics{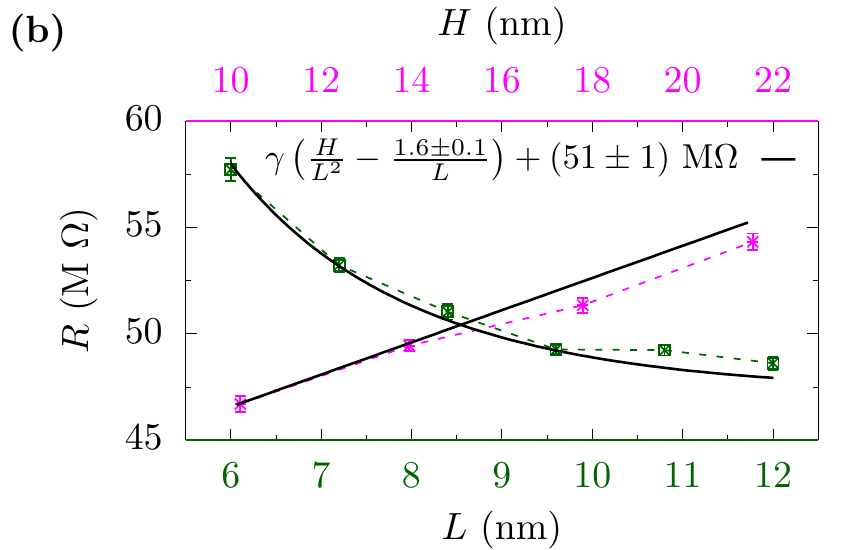}
\caption{\label{Fig2} {\bf Electric potential and total resistance.} (a) The potential $V$ (color map with contour lines at 0.05 V intervals) and normalized current density (arrows) with a 1 V applied potential and a 1.18 nm pore radius. The resistance is large in the pore region, resulting in a large electric field across the membrane [over about 1 nm]. 
(b) Resistance versus the cell height $H$ (magenta) and cross-sectional length $L$ (green) for the pore in (a). For $R$ versus $H$ ($L$), we use $L=9.6$ nm ($H \approx 14$ nm). Equation~\eqref{model} provides a good fit to the data, yet it predicts $R_\infty$ is higher than where the data apparently converges. This is due to the bulk dimensions not changing in tandem as the ansatz indicates should be done. The value of $R_\infty$, though, is consistent with the proper scaling procedure, see Figure~\ref{Fig4}. We use the electrolyte resistivity from MD simulations of a bulk-only cell, which gives $\gamma\approx 70$ M$\Omega$ nm. Unless otherwise noted, all error bars are $\pm 1$ block standard error.}
\end{figure*}

Despite the intense and broad interest in ion transport, one of its most fundamental aspects, the convergence of the bulk to the pore, is essentially not computable with all-atom molecular dynamics (MD)~\cite{yoo2015}, yet is very important for understanding {\it in vivo} operation and characteristics of ion channels~\cite{alcaraz2017ion}. Experiments on mono- or bi-layer graphene, show a dominant $1/a$ access resistance for a pore of radius $a$ \cite{garaj2010,garaj2013,schneider2013} as expected for an atomically thin pore. Other experiments, however, seemingly yield $1/a^2$ behavior~\cite{Schneider2010}.  Moreover, simulations give contradictory results, some \cite{Hu2012} with $1/a$ and others \cite{sathe2011} $1/a^2$. We develop a finite-size scaling analysis for all-atom MD to extract the full resistance, both access and pore, to allow direct comparison with experimental results. Using this, we show that graphene pores, see Figure~\ref{Fig1}, have both an access and pore resistance contribution all the way to the dehydration limit.

Hall's form of access resistance \cite{Hall1975}  is the classic result for ions to converge from bulk, far away from the pore, to the pore mouth, 
\begin{equation}\label{Hall}
\RH = \frac{\gamma}{4  a} ,
\end{equation} 
where $\gamma$ is the electrolyte resistivity and $a$ is the pore radius. When taking this resistance for both sides of the membrane, it is the same form of resistance originally given by Maxwell \cite{maxwell_1881} and later by Holm \cite{Holm1958contact} and Newmann \cite{newman1966}  for the {\em electrical}  ``contact'' resistance of a circular orifice, which has a ballistic counterpart known as the Sharvin resistance~\cite{Sharvin65-1}. Maxwell's formula for contact resistance is valid when the radius of the orifice is much larger than the mean free path of the electrons but in general the electric contact resistance is a combination of the Maxwell and Sharvin resistance~\cite{wexler1966,nikolic_1999}.The access resistance for ion transport, however, does not have any ballistic component. We also note that the same form of access resistance is also present in thermal transport~\cite{GrayMathews1895,grober1921} and gas diffusion~\cite{brown1900}. 

The above result assumes a hemispherical symmetry and homogeneous medium (that is, no concentration gradients, even near the pore, and no charges or dipoles on the membrane), as well as an infinite distance between the pore and electrode.  These assumptions can hold for small voltages and for well-fabricated pores (for example, recent low-aspect ratio pores show only an access contribution following Eq.~\eqref{Hall} \cite{tsutsui2012}). Moreover, factors such as surface charges~\cite{aguilella2005}, concentration gradients~\cite{luchinsky2009Self,peskoff1988electrodiffusion}, and an asymmetrical  electrolyte~\cite{lauger1976} will influence the access resistance.

Hall's form of access resistance is independent of bulk size, which will hold so long as the bulk dimensions are large and balanced (that is, the height of the cell should not be disproportionately large compared to its cross-sectional length). In confined geometries, however, strong boundary effects or unbalanced dimensions modify this behavior (for example, in scanning ion conductance microscopy the imposed boundary close to the pore causes the access resistance to deviate from Eq.~\eqref{Hall}~\cite{korchev1997,panday2016}). In MD, in particular, the simulation cells are both highly confined and periodic to collect sufficient statistical information on ion crossings.
We thus examine the access resistance for a finite bulk. Its derivation is easier in rotational elliptic coordinates \cite{maxwell_1954,Holm1958contact,newman1966,braunovic2006}, $\xi$ and $\eta$, which are related to cylindrical coordinates, $z$ and $\rho$, via
\begin{align}
z & =  a \xi \eta  \\
\rho & =  a \sqrt{(1+\xi^2)(1-\eta^2)} .
\end{align}
Laplace's equation for the potential then becomes
\begin{equation}\label{laplace}
\frac{\partial }{\partial \xi} \left[ (1+\xi^2) \frac{\partial V}{\partial \xi} \right]+\frac{\partial }{\partial \eta} \left[ (1-\eta^2) \frac{\partial V}{\partial \eta} \right]=0 .
\end{equation}
For boundary conditions, we consider a spheroidal electrode, representing the equipotential surfaces that form even when a flat electrode is present, and a circular pore. That is, (1) $V=0$ on the pore mouth ($\xi=0$), (2) $V=V_0$ on a spheroidal electrode at distance $l$ ($\xi=l/a$), and (3) $\partial V/ \partial \eta= 0$ on the membrane surface ($\eta=0$).

Although clearly idealizations, we see features that reflect these boundary conditions from all-atom MD. Applying a constant electric field along the $z$-axis gives rise to the ion flow patterns and electric fields in Figure~\ref{Fig2}. Due to the pore resistance, a charged double layer forms \cite{grahame1947}, with enhanced cation (anion) density on the positive (negative) voltage side. The potential at the pore mouth (which is essentially the whole pore due to the atomic thickness) is not constant, but is roughly so. The deviation is mainly due to the potassium ions coming closer to the membrane than chloride ions, pushing the potential outward. That is, the asymmetry between cations and anions (in sizes, charges, interactions), as well as other effects, distort the potential surface. The equipotential surfaces have roughly a spheroidal form (with deviation due to both simulation error, the accumulated simulation time needs to be very large, and also due to atomic-scale features of the graphene, water, and ions). Due to the large voltage and the non-zero pore resistance, only boundary condition (3) does not appear to be present. However, we expect the right functional dependence of the finite-size deviation from the Maxwell-Hall form.

Using those boundary conditions, Eq.~\eqref{laplace} yields 
\begin{equation}
 \frac{V}{V_0} =  \frac{\tan^{-1}\xi}{\tan^{-1} (l/a)} .
\end{equation} 
The ionic current through the pore is then
\begin{equation}
I = \frac{1}{\gamma} \int_0^{a} \frac{\partial V}{\partial z}\Big|_{z=0}\ \rho d\rho =\frac{2 \pi a V_o}{\gamma\tan^{-1} (l/a)} ,
\end{equation}
giving the access resistance 
\begin{align}\label{access}
R_\mathrm{access} &= \frac{\gamma \tan^{-1}(l/a)}{2 \pi a} \approx \RH \left(1 - \frac{2 a}{\pi l}\right),
\end{align} 
where the approximation is up to $\mathcal{O}\left[\frac{\rp}{l}\right]^3$ (when $l$ is about 2$a$, the higher order corrections are small, about 2.6 \%, likely much smaller than corrections due to atomic details at this scale).
In confined geometries, one needs to account for $\rp/l$ correction term, especially in MD where the computational cost typically keeps the ``bulk'' dimensions around 10 nm.

\begin{figure}[t]
\includegraphics[width=\columnwidth]{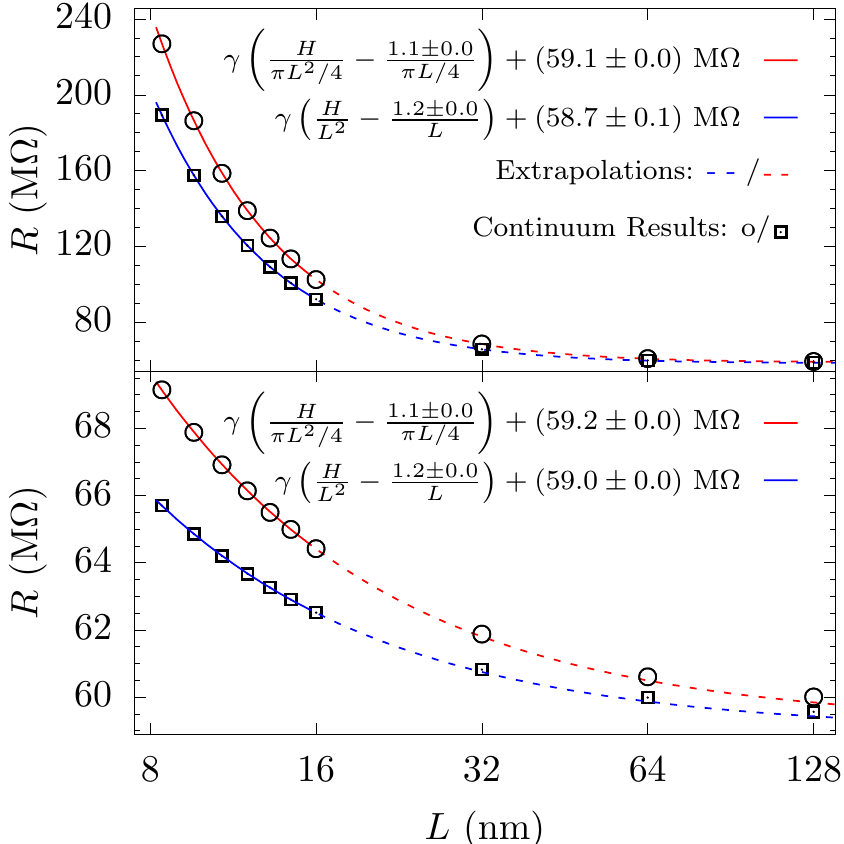}
\caption{\label{Fig3} {\bf Pore and access resistance in continuum simulations.} Resistance versus the bulk dimension $L$ for cylindrical (circles) and rectangular (squares) cells. The membrane height is 1 nm (that is, approximately that for graphene plus the charged double layer) and the pore radius is $a = 1$ nm (and $\gamma$ is from MD for consistency). The bulk height is fixed to $H=140$ nm in the top panel and the aspect ratio is fixed at $H=2L$ in the bottom panel. We fit Eq.~\eqref{model} for $L\le 16$ (solid lines) and extrapolate to larger  $L$ (dashed lines). The fit accurately determines $R_\infty$. Hence, the small simulation sizes in all-atom MD should be sufficient to obtain $R_\infty$. }
\end{figure}

Away from the membrane, the equipotential surfaces start to become flatter, taking on a bulk-like form. That is, the flow lines, while pointing towards the pore near its entrance/exit, orient along the $z$-axis further away, as do the electric field lines. For a simulation cross-sectional area of $A=\mathcal{G} L^2$, where $\mathcal{G}=\pi/4$ for a cylindrical cell and $\mathcal{G}=1$ for a rectangular cell, the access region must end by $l = f_1 L/2$, with $f_1 \sim \mathcal{O}(1)$, as the ellipsoidal potential surfaces encounter the cell boundary. Sometime afterward, at $f_2 L/2$ with $f_2 \sim \mathcal{O}(1)$, a normal bulk region appears. Thus, the total resistance is approximately 
\begin{align}\label{accessFinite}
R = & 2 \left(\RH - \frac{\gamma}{2 \pi  f_1 L/2}  \right) + \frac{\gamma (H-2 f_2 L/2)}{\mathcal{G}L^2}  + \frac{\gamma f_3}{\mathcal{G}L}+ R_\mathrm{pore} .
\end{align} 
The first (access-like) term occurs on both sides of the membrane (giving the factor of 2). The second (bulk-like) term uses the total height $H$ minus the two access/transitory regions of height $f_2 L/2$ ($H$ does not include the membrane thickness and charged double layers, and it must be reasonably larger than $f_2 L$). Figure~\ref{Fig2}(b) shows we indeed have this bulk-like region as the resistance increases linearly with $H$. The third term is a correction, $\gamma f_3/\mathcal{G}L$, to account for the resistance of the transition region between the access and the normal bulk, both of which would drop as $1/L$ in that finite region.

We note that some previous studies have shown the dependence of the ionic current on the cell height~\cite{Gumbart2012,jensen2013}. However, in Ref.~\onlinecite{Gumbart2012}, the dependence is examined in the context of changing field with the height and, in Ref.~\onlinecite{jensen2013}, the difference is considered insignificant. In linear response, the pore resistance should be independent of the applied field. While we have a 1 V potential, the main findings hold for smaller voltages, as continuum simulations demonstrate, and there is roughly linear behavior of the graphene I-V curve at this voltage~\cite{Sahu2017NanoLett}.

Since all three corrections depend on $1/L$, we can combine them into a single term, yielding
\begin{align}\label{model}
R =& \gamma\left( \frac{H}{\mathcal{G} L^2}- \frac{f}{\mathcal{G} L}\right)+R_\infty ,
\end{align}
where $R_\infty$ is the combined access and pore resistance when all the linear dimensions of the cell are balanced and large compare to the  pore radius. The behavior of $R_\infty$ is expected to be $R_\infty=2\RH+R_\mathrm{pore}$ from Hall's theroy, which we will show later to hold for graphene pores down to the dehydration limit. The factor $f=2 \mathcal{G}/\pi  f_1 + f_2 - f_3$ depends on geometric details of the cell. Assuming $f_1 \approx f_2 \approx 1$ (and $f_3$ small), $f \approx 1.6$ for a rectangular and $f \approx 1.5$ for a cylindrical cross-section. The estimates will remain close even if $f_3$ is substantial, so long as the transitory region is approximately a mix of access and bulk-like behavior. Despite these estimates, we treat $R_\infty$ and $f$ as fitting parameters. 

Figure~\ref{Fig2}(b) already shows that this scaling form can capture the dependence of the resistance on the cell dimensions. However, a very peculiar behavior arises: $R_\infty$ is above the decay of $R$ with $L$. The scaling form, though, suggests that one should take $H=\alpha L$, where $\alpha$ is the cell aspect ratio, reducing Eq.~\ref{model} to $R =\gamma\left( \frac{\alpha-f}{\mathcal{G} L}\right)+R_\infty$. This indicates that if we knew $f$ exactly, we could take $\alpha=f$, that is, a ``golden aspect ratio'' (the estimated $f$ is not the actual golden ratio, $(1+\sqrt{5})/2$) to remove the $L$-dependence of $R$ and obtain $R=R_\infty$ for a finite size simulation cell. Of course, if the simulation cell is too small, the potential and densities will be artificially distorted at the periodic boundary (or finite edge). Since we do not know $f$ exactly, we will take $\alpha=2$, somewhat larger than the expected value of $f$, which will simultaneously ensure that $R$ converges to $R_\infty$ from above and reduce the amount that $R$ changes as $L$ increases. As well, $H$ should be reasonably larger than twice the access region, as otherwise ions would have unusual flow patterns. We prove the existence of the golden aspect ratio using continuum simulations in Ref.~\cite{sahu2017golden}.

We first examine Eq.~\eqref{model} with continuum simulations, that is, using Laplace's equation, of both rectangular and cylindrical (finite) cells using a commercial finite element solver. Figure~\ref{Fig3} shows that continuum simulations yield good agreement with the ansatz and allow for the extrapolation of $R_\infty$ using small simulation cells, which bodes well for the small simulation sizes typical of all-atom MD. Moreover, it suggests that using the constant aspect ratio cells is better, as it yields less deviation over all.

\begin{figure}[t]
\includegraphics[width=\columnwidth]{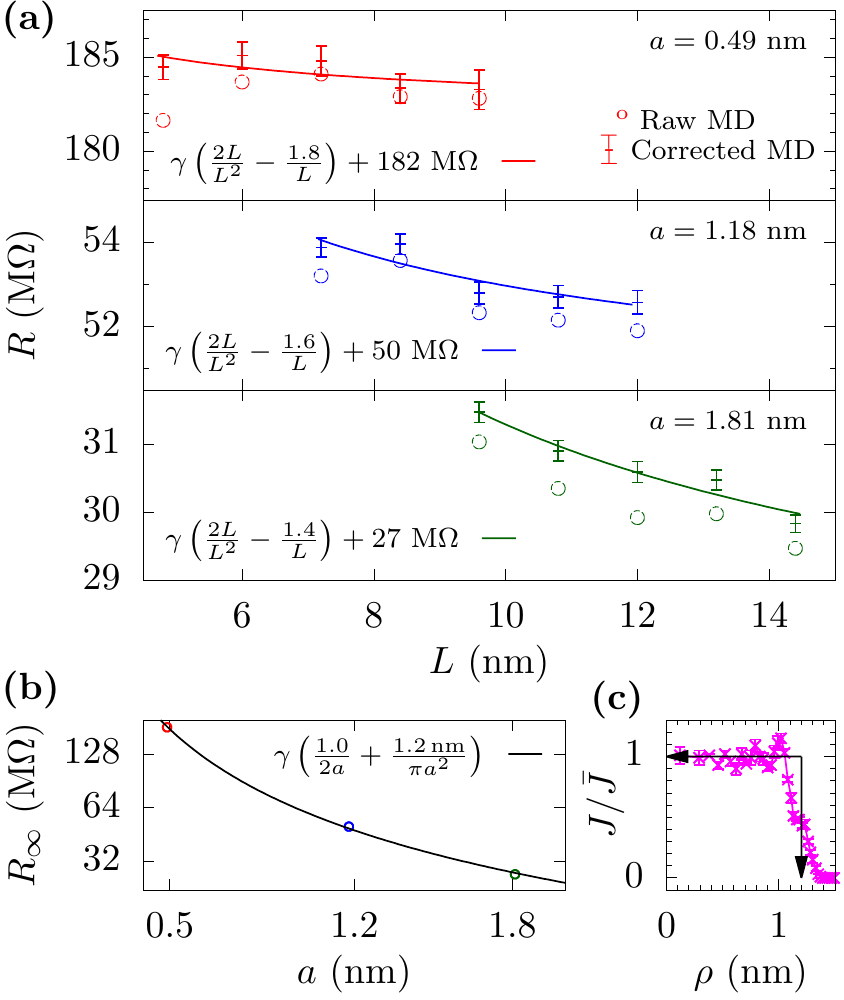}
\caption{\label{Fig4} {\bf Pore and access resistance in graphene}. (a) The three panels show the resistance versus cross-sectional length of the simulation cell for different pore sizes. The resistance from MD is shown in circles and the resistance corrected to $H=2L$ is used for fitting the model (equilibration changes $H$ from its initial value). The scaling analysis suggests the ``golden aspect ratio'' of $H/L \approx (1+2/\pi)$ will remove $L$ dependence of $R$. We choose an aspect ratio slightly above this so that the variation of $R$ is small but that the infinite system limit is approached from above. Unbalanced cells, for example, Figure~\ref{Fig2}(b) and Figure~\ref{Fig3} upper panel, give much larger changes in $R$ and can also result in unusual convergence to $R_\infty$. (b) The extracted $R_\infty$ versus the pore radius $a$ indicates that there is a Maxwell-Hall access contribution. (c) Normalized current density inside a pore ($a=1.18$ nm) showing that the effective pore radius is about $0.25$ nm smaller than the geometric radius. The errors in $f$ and $R_\infty$ are approximately $\pm 0.1$ and $\pm 1$ M$\Omega$, respectively, for all of the pores.} 
\end{figure}

We now employ our finite-size scaling ansatz to examine the total resistance in graphene nanopores. Figure~\ref{Fig4}(a) shows the resistance versus $L$ for $H=2L$. Using the extracted $R_\infty$, we can determine the behavior of the resistance versus $a$ (due to computational cost, we examine only a small range of $a$), see Figure~\ref{Fig4}(b). We find that even at the nanometer scale, the resistance of graphene follows the continuum form
\begin{align} \label{measured}
R_\infty &= 2\RH+R_\mathrm{pore} =\frac{\gamma}{2 a}+\frac{\gamma h_p}{\pi  a^2} .
\end{align}
However, the radius can not be taken as the geometric radius (the largest circle that will fit within the pore, even correcting for van der Waals interactions). Rather, the radius is determined by the accessible area in the pore. Figure~\ref{Fig4}(c) shows how the current density in the pore tapers off as the radial coordinate increases (see also the SI). Hence, taking the pore radius from the actual effective area for current to flow accounts for hydration layers around the ions and van der Waals interactions, as well as fluctuations of the pore edge. Doing so, we find $R_\infty = \gamma  /2 a +h_p^{eff} \gamma/\pi a^2$ with $h_p^{eff}=(1.2\pm0.1)$ nm. That is, we find the Maxwell-Hall access contribution and an effective thickness of 1.2 nm, in agreement with the charged double layer separation. This thickness is larger, but within the error, of the 0.6 nm value found experimentally~\cite{garaj2010,schneider2013}, where, however, the voltage was an order of magnitude smaller and thus the charge double layer was less prominent. 

Thus, the resistance is a combination of both $1/a$ and $1/a^2$ behavior. Contextual aspects due to, e.g., van der Waals interactions, hydration layers, edge fluctuations, charge double layers, and potentially effective ion mobilities in the pore, obscure the parameters that appear in $R_\infty$, making it difficult to determine the dependence of the resistance on the radius. Indeed, the proper pore radius, the one related to the accessible area, is crucial. Experimentally, there are many sources of ambiguity: Uncertainties in measured values and in the pore depth (for example, multi-layer versus single layer graphene) and pore size (and aspect ratio / non-circularity), plus unknown charged functional groups or dipoles (that would enhance $1/a$ behavior by creating excess density at the membrane surface that ``feeds'' the current through the pore via its circumference), all affect either the balance of $1/a$ and $1/a^2$ behavior, or how well one can extract that behavior. This list can also include nonlinearities (for example, MD simulations show the onset of polarization-induced chaperoning of ions~\cite{Sahu2017NanoLett}, which can tilt the balance in favor of access resistance as the dominant resistance). Different membranes and conditions can thus display diverse behavior, but ``ideal'' graphene membranes with pores larger than the dehydration limit have both access and pore contributions. As the pore radius increases, though, access resistance will dominate, as seen in Ref.~\onlinecite{garaj2010}. The observation of $1/a^2$ behavior must be due to interpretation (for example, the inclusion of multi-layer membranes in the data fitting, or the fitting itself) or to some unknown aspect of the experimental setup. 

Our results demonstrate that one can capture pore and convergence resistance in reasonably sized simulations, despite the long-range nature of the access resistance. 
One may also extract separately the access and pore contributions to resistance, which, however, would require knowing where to partition the voltage drop (in the presence of charge double layers and other nanoscale structure, this is not a simple task).
Thus, when designing porous membranes, one can use MD to both capture the ``contextual'' aspects of the pores, atomic scale details such as charges, fluctuations, and geometry, and the influence of the bulk electrolyte. This will allow for a quantitative comparison between measurements and simulations. Moreover, filtration and other nanopore technologies typically require many pores. The access contribution in such porous membranes is crucial, as it can undergo a transition into collective behavior when the pore density is high. Inevitably, there will be a trade off between the physical dimensions of these simulations and the time scales (and voltages) reachable. Our finite-size scaling ansatz, Eq.~\eqref{model}, gives a theoretical approach to guide this trade off and determine the influence of convergence. \\
 
\section*{Methods}
We used NAMD2 \cite {phillips2005} to perform all-atom molecular dynamics simulations with 2 fs integration time step and periodic boundary condition in all direction.  The force field parameters is rigid TIP3P \cite{jorgensen1983} for water and from CHARMM27 \cite{Feller2000} for the rest of the atoms. Short range electrostatic and van der Waals forces have cutoff of 1.2 nm. However, full electrostatic calculation occur every 4 time steps using the Particle Mesh Ewald (PME) method. 

\section*{ACKNOWLEDGMENTS}
We thank S. Stavis for helpful discussions. S. S. acknowledges support under the Cooperative Research Agreement between the University of Maryland and the National Institute of Standards and Technology Center for Nanoscale Science and Technology, Award 70NANB14H209, through the University of Maryland.

\bibliography{./reference} 
\end{document}


\title{Maxwell-Hall access resistance in graphene nanopores -- Supplementary Information}
\author{Subin Sahu}
\affiliation{Center for Nanoscale Science and Technology, National Institute of Standards and Technology, Gaithersburg, MD 20899, USA}
\affiliation{Maryland NanoCenter, University of Maryland, College Park, MD 20742, USA}
\affiliation{Department of Physics, Oregon State University, Corvallis, OR 97331, USA}

\author{Michael Zwolak}
\email{mpz@nist.gov}
\affiliation{Center for Nanoscale Science and Technology, National Institute of Standards and Technology, Gaithersburg, MD 20899, USA}

\maketitle
\tableofcontents
\clearpage

\section{Methods}

\subsection{All-atom MD simulations}

Our system consists of a single-layer graphene membrane with a pore in the center and in 1 Mol/L KCl solution, as shown in Figure 1 of the main text. We build the system using VMD 1.9.1 \cite{humphrey1996vmd} and perform all-atom molecular dynamics simulations using NAMD2 \cite {phillips2005} with periodic boundary condition in all directions. The force field parameters are rigid TIP3P \cite{jorgensen1983} for water and CHARMM27 \cite{Feller2000} for the other atoms. We fix the outer edge of the graphene membrane but the bulk of the membrane has no confinement other than the C-C bonds of graphene. The simulations have an integration time step of 2 fs and Langevin damping of 0.2 ps for only carbon and water (via its oxygen atoms). Non-bonded interactions (van der Waals and electrostatics) have a cutoff of 1.2 nm. However, full electrostatic calculations occur every 4 time steps using the Particle Mesh Ewald (PME) method. We first minimize the energy of the system for 4000 steps (8 ps) and then heat it to 295 K in another 8 ps. A 1 ns NPT (constant number of particles, pressure and temperature) equilibration using the Nose-Hoover Langevin piston method \cite{Martyna1994} -- to raise the pressure to 101~325 Pa (i.e., 1 atm) -- followed by 3 ns of NVT (constant number of particles, volume and temperature) equilibration generates the initial atomic configuration. An electric field perpendicular to the plane of the membrane (1 V potential difference) drives the ionic current through the pore.\\

\begin{figure}[h]
\includegraphics{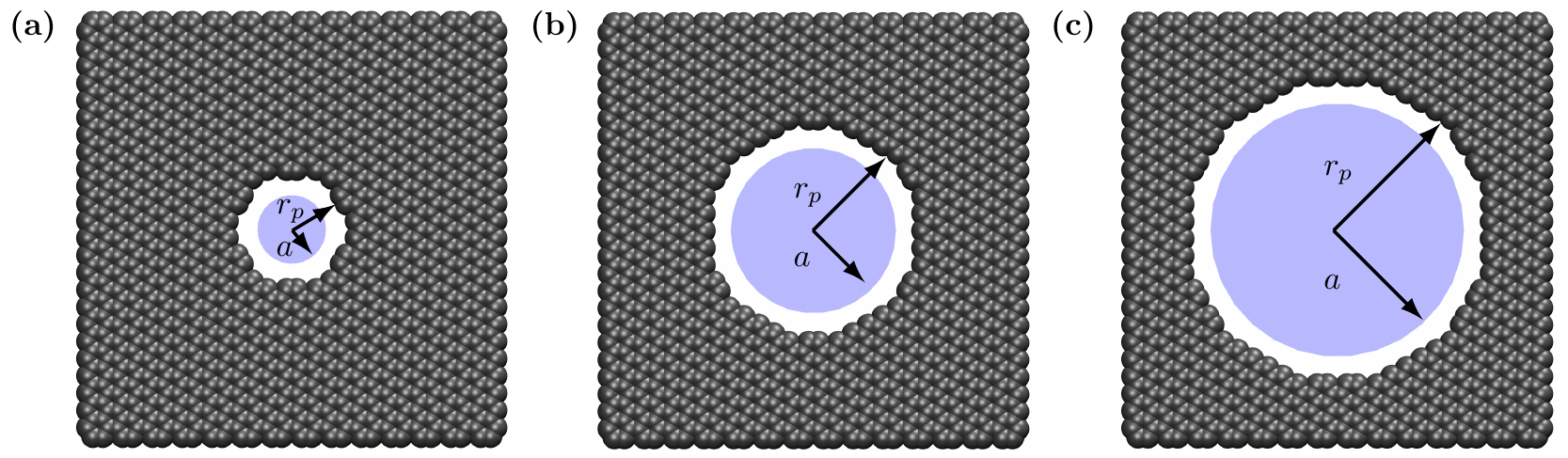}
\caption{\label{pores} A 6 nm $\times$ 6 nm section of the graphene sheet showing the pores with effective radii (a) $\rp=\ra$ nm,  (b) $\rp=\rb$ nm, and (c) $\rp=\rc$ nm. We construct the pores by removing all carbon atoms within 0.8 nm, 1.5 nm, and 2.2 nm, respectively, of the pore center and also removing any dangling bonds. The geometric definition of the pore radius, $r_p$, is the average distance between the center of the pore and the inner edge of the pore atoms (i.e., carbons with a size given by their vdW radius). However, the effective pore radius, $a$, is about 0.2 nm smaller than $r_p$ due to the finite size (hydration and vdW radii) of the ions. The schematic view here is in agreement with the statistical view of ion crossings, see Figure~\ref{radius}, with the exception of some minor contextual issues arising from the pore atomic structure (e.g., the ion crossings have a clear hexagonal symmetry).}
\end{figure}

\begin{figure}[h]
\includegraphics{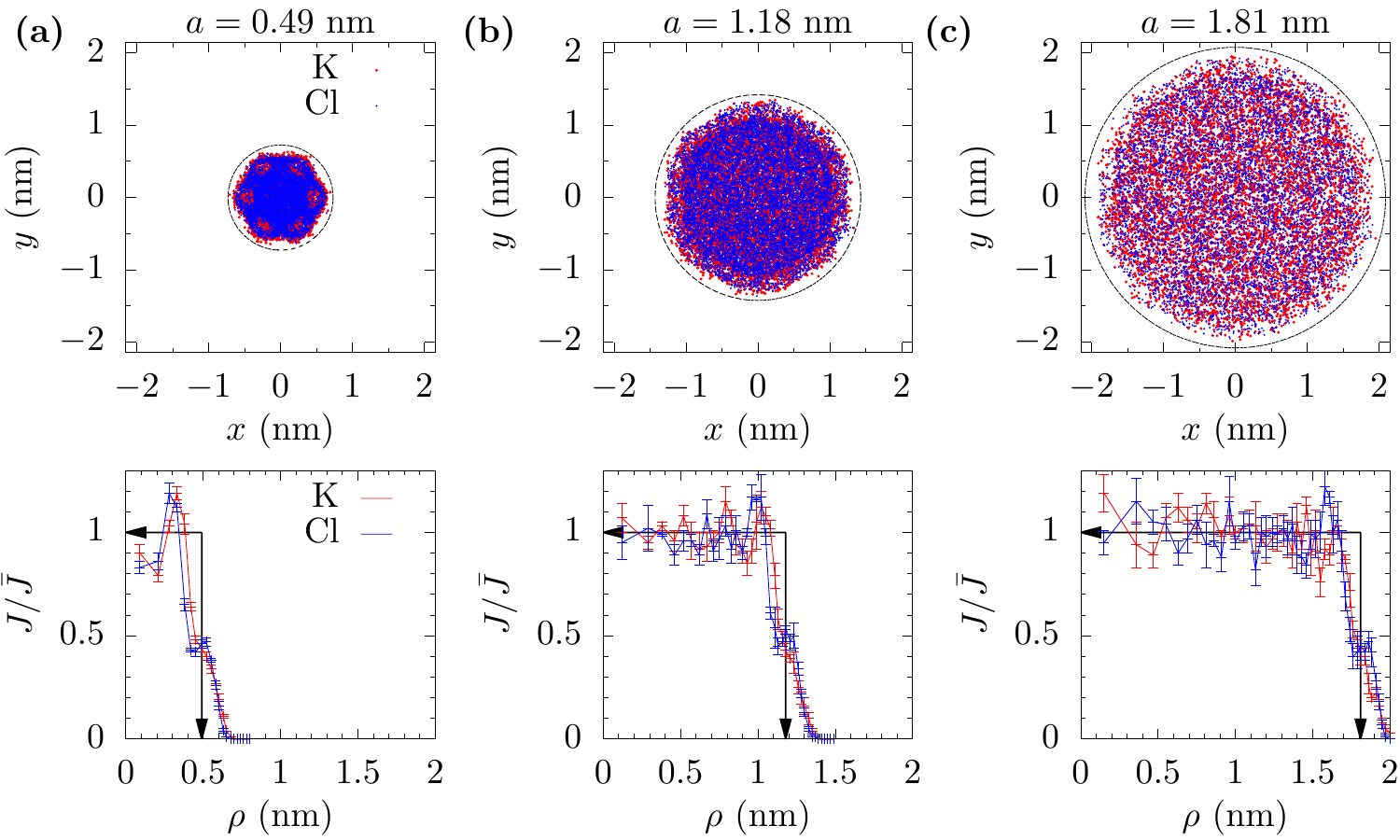}
\caption{\label{radius} Current density of K$^+$ and Cl$^-$ crossing pores with radii (a) $\rp=\ra$ nm, (b) $\rp=\rb$ nm, and (c) $\rp=\rc$ versus the cylindrical coordinate $\rho=\sqrt{x^2+y^2}$. The top panels show scatter plots of ion crossings in the $xy$-plane. The circles indicate $r_p$ -- the geometric pore radius -- which is about 0.2 nm bigger than the radius of the accessible area, as seen by the gap in the ion crossing events. The bottom panels show variation of current density $J$, normalized by ``unattenuated'' current density $\bar{J}$, with radial coordinate $\rho$ inside the pore. The black arrows show the effective pore radius from this distribution of $J$. Error bars are $\pm 1$ standard error from six parallel simulations.}
\end{figure}

\subsection{Pore radius}
We consider three pore sizes with effective radius $\rp=\ra$ nm, $\rp=\rb$ nm, and $\rp=\rc$ nm, as shown in Figure \ref{pores}. Geometrically, the radius of the pore can be defined as the average distance between the center of the pore and the pore atoms at the edge minus the van der Waals radius of carbon (0.17 nm), i.e., the average distance between the center of pore and edge of the carbon atoms. However, the radius of the accessible area for the transport of ions is about 0.2 nm smaller than $r_p$, as shown in Figure \ref{radius}. The exclusion near the pore edge is due to van der Waals (vdW) repulsion (i.e., the finite ion size since we already account for carbon's vdW radius) and dehydration. Thus, we define the effective radius, $\rp$ of the pore as 
\begin{equation} \label{eq:effrad}
\pi a^2 \bar{J} =\int_0^{r_p} J(\rho)2\pi \rho d\rho ,
\end{equation}
where $J(\rho)$ is the current density at radial coordinate $\rho$ (assuming cylindrical symmetry, which is reasonable for graphene pores but not perfect -- relaxing this would require much longer simulations to acquire sufficient statistics on the angular dependence of ion crossings) and $\bar{J}$ is the average current density in the region of the pore where $J(\rho)$ is flat. This calculation is essentially weighing the area contributions by the Boltzmann factors at that location, except we use the out-of-equilibrium probability distribution of ion crossing events instead of the Boltzmann factors from the free energy barriers. The quantity $\bar{J}$ serves the role of an ``unattenuated'' current density -- i.e., the current density where there is no excess free energy barrier. We note that fluctuations of the graphene membrane, specifically around the pore edge, also affects the pore size and its effect is included in Eq.~\eqref{eq:effrad}.

 \subsection{Error analysis for convergence in time}

We compute the error in the MD results using the block standard error (BSE) method \cite{grossfield2009}. We divide a single MD run of duration $T$ into number of contiguous blocks of equal duration $\tau$. The BSE is given by 
\begin{equation}
\mathrm{BSE} = \frac{s_\tau \sqrt{\tau}}{\sqrt{T}},
\end{equation}
where $s_\tau = \sqrt{ \frac{\sum_i (\avg{I_\tau}_i -\avg{I_T})^2}{(N_b -1)}} $  is the standard deviation of the mean current $\avg{I_\tau}$,  within each of the $N_b$ blocks. The error bars in the plots are $\pm 1$ BSE unless otherwise noted.

\section{Finite-size scaling}

As mentioned in the main text, if all the linear dimensions of the cell (experimental or theoretical) are simultaneously taken to be large, the normal bulk component of the resistance will vanish and the measured resistance is expected to take on the form
\begin{align} \label{measured}
R_\infty &=\frac{\gamma}{2 \rp}+\frac{\gamma h_p}{\pi  \rp^2}
\end{align}
in the continuum limit and assuming a cylindrical pore of height $h_p$. $R_\infty$ can be found by using the finite-size scaling
\begin{align}\label{modelMD}
R =& {\gamma}\left( \frac{H}{\qg L^2}- \frac{f}{\qg L}\right)+R_\infty,
\end{align} 
where $\qg L^2$ is the cross-sectional area of the cell, $L$ is the cross-sectional length, $\qg$ is a geometric factor ($\mathcal{G}=\pi/4$ for a cylindrical cell and $\mathcal{G}=1$ for a rectangular cell), and $f$ is the fitting parameter.

\begin{figure}[t]
\includegraphics {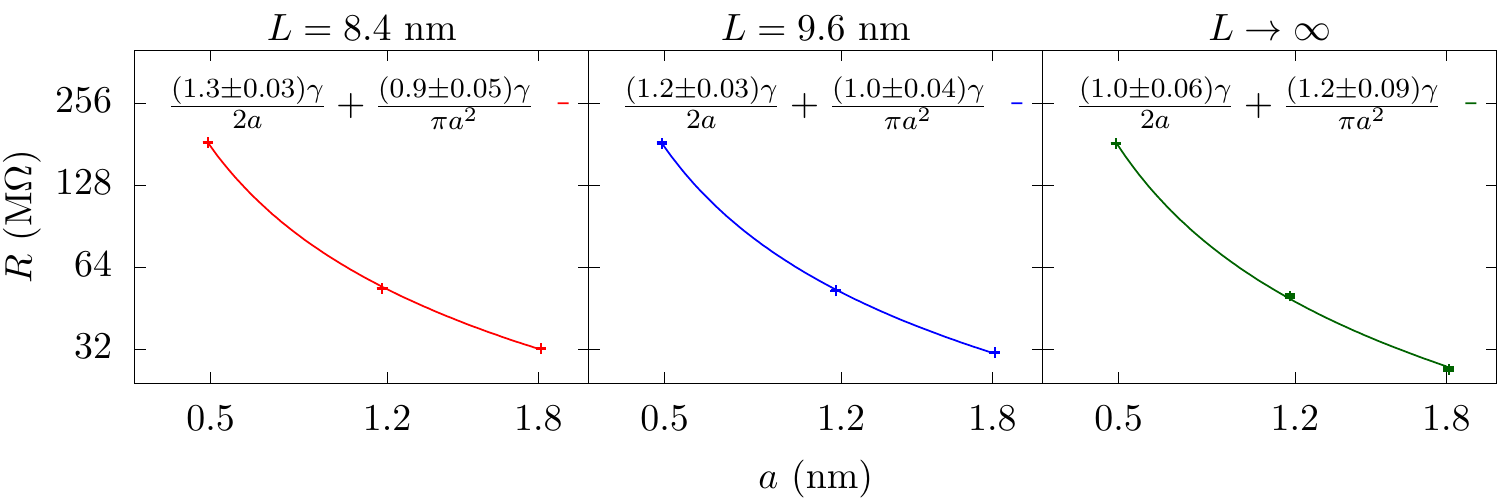}
\caption{\label{fits} Maxwell-Hall form of the access resistance fitted for $L=8.4$ nm, $L=9.6$ nm, and $L\rightarrow \infty$ (extrapolated resistance). Only when $L\rightarrow \infty$ does the exact Maxwell-Hall form emerge for the access resistance, i.e., a coefficient of 1 times $\gamma/2a$. The error bars are $\pm 1$ BSE.}
\end{figure}

\begin{figure}[t]
\includegraphics{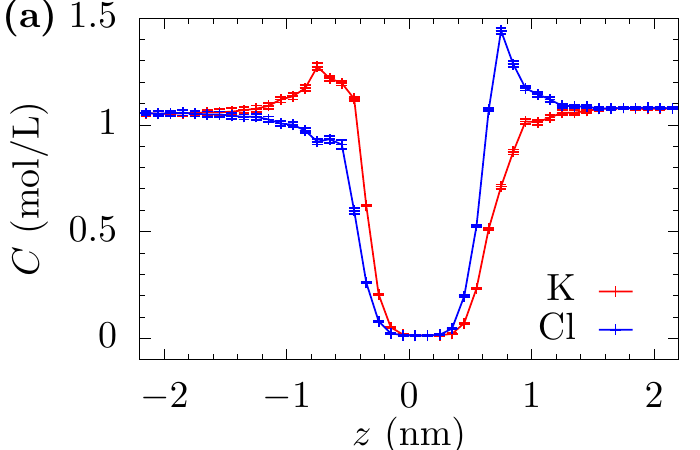} \quad
\includegraphics{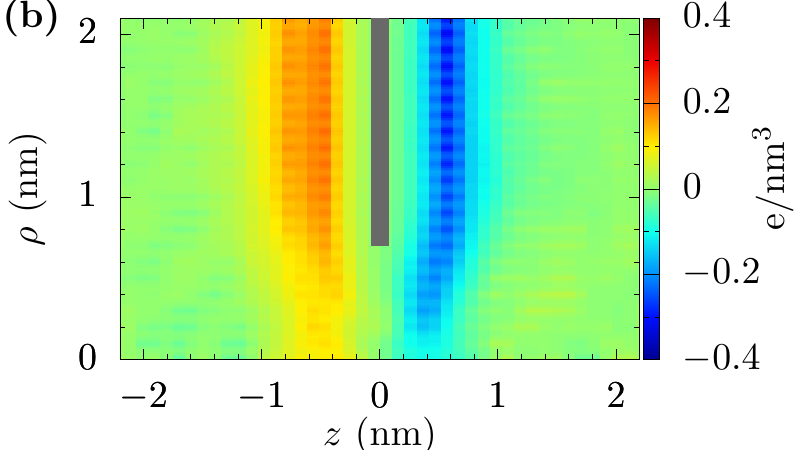} 
\caption{\label{ionDen} (a) $\Ki$ and $\Cli$ ion densities, and (b) net charge density when a 1 V voltage is applied across a graphene membrane with a radius $\rp=\ra$ nm pore. The peak of $\Ki$ ion concentration is closer to the graphene membrane than that of $\Cli$ due to smaller ionic size of the former. The $z$ distance between the two peaks is about $1.5$ nm. The error bars are $\pm 1$ standard error from six parallel simulations.} 
\end{figure}

In Figure~\ref{fits}, we fit a modified form of Eq.~\ref{measured} for $L=8.4$ nm, $L=9.6$ nm, and $L\rightarrow \infty$, 
\begin{align} 
R_L &=\frac{b_L \gamma}{ 2 \rp}+\frac{h_L \gamma}{\pi  \rp^2} ,
\end{align}
where $b_L$ and $h_L$ are the fitting parameters.  For $L=8.4$ nm and $L=9.6$ nm, the access resistance is larger than the Maxwell-Hall form due to the unbalanced dimensions of the cell and the cell's relative size compared to the differing pore radii. Only when $L\rightarrow \infty$ do we get exactly the Maxwell-Hall value. Also, the fitted value of the membrane thickness is $\approx 1.2$ nm. This is consistent with the separation of ion density peak on the two sides of the graphene membrane (i.e., the charge dipole layer separation), as seen in Figure~\ref{ionDen}.

\section{Electric fields and current density}

We calculate the electrostatic potential and the charge density using the VolMap plugin of VMD. The current density is the average ion displacement between the snapshots (10 ps) over the length of the simulation,
\begin{align}
\vec{J}(\vec{r})&=\frac{\sum_i q_i \vec{v}_i(\vec{r})}{dV},
\end{align}
where the sum is over all the ions within the volume element $dV$(with $dx=dy=dz=0.1$ nm) at position $\vec{r}$.

\begin{figure}[t]
\includegraphics {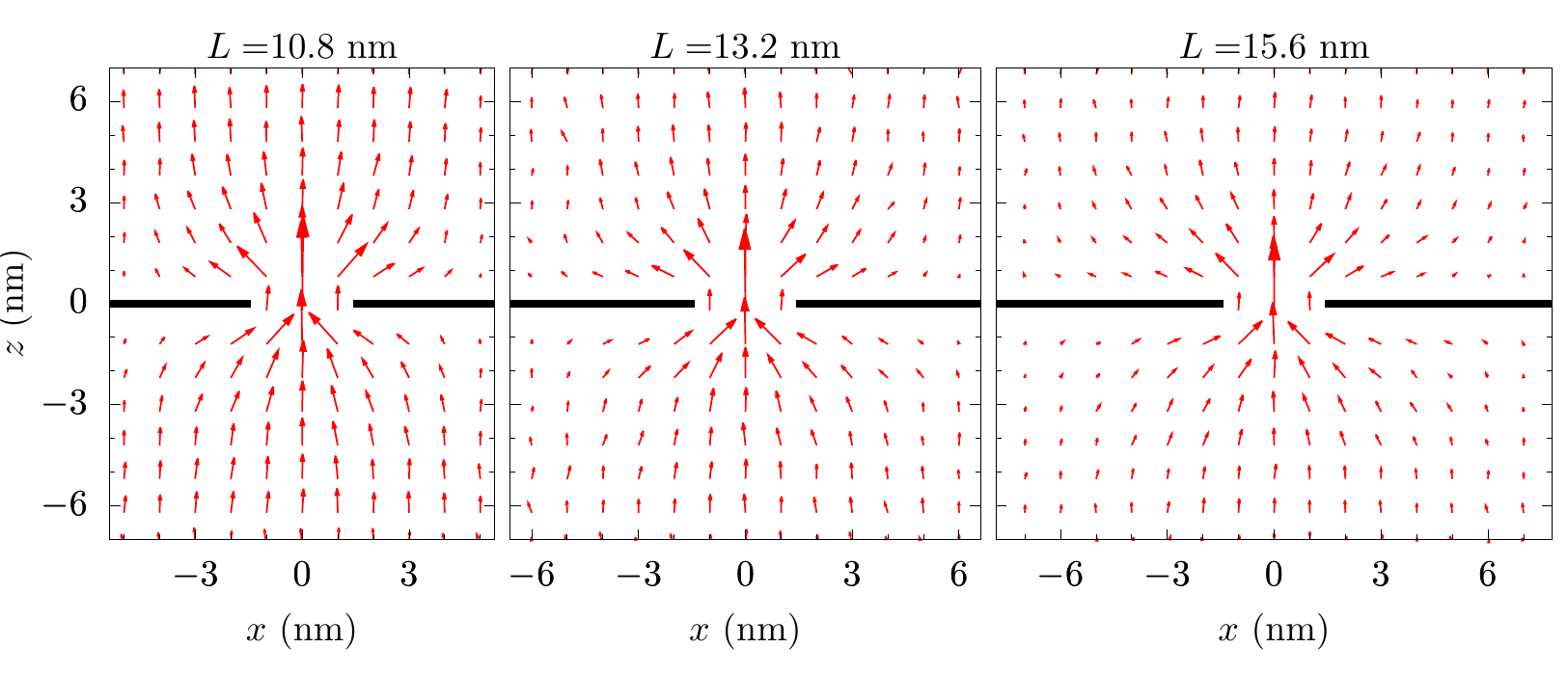}
\caption{\label{Js} The current density, $J$, showing the flow pattern for different cell cross sections from MD simulations of a graphene nanopore with radius $\rp=\rb $ nm. The flow quickly orients with the $z$-axis away from the pore regardless of the cross-sectional area of the cell. Note that $J$ is not constant with $L$ due to a changing balance of bulk and pore resistance.}
\end{figure}

Figure \ref{Js} shows the flow pattern for three different cell cross sections with a pore radius $\rp=\rb$ nm. In each of them, the current density $J$ quickly orients along $z$-axis. It is also seen from Fig \ref{Js} that $J$ decreases with $A=\qg L^2$, which can be understood by looking at the average value of $J_z$ according to our model,
\begin{align}
\langle J_z \rangle = \frac{V}{R A}=\frac{V}{\gamma(H-f L)+R_\infty \qg L^2}.
\end{align}

\section{Bulk resistivity}

\begin{figure}
\includegraphics[width=.48\textwidth]{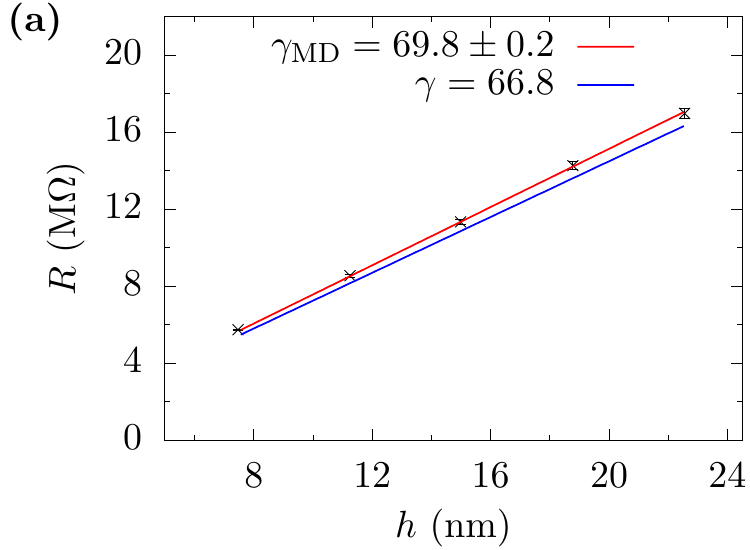}
\includegraphics[width=.48\textwidth]{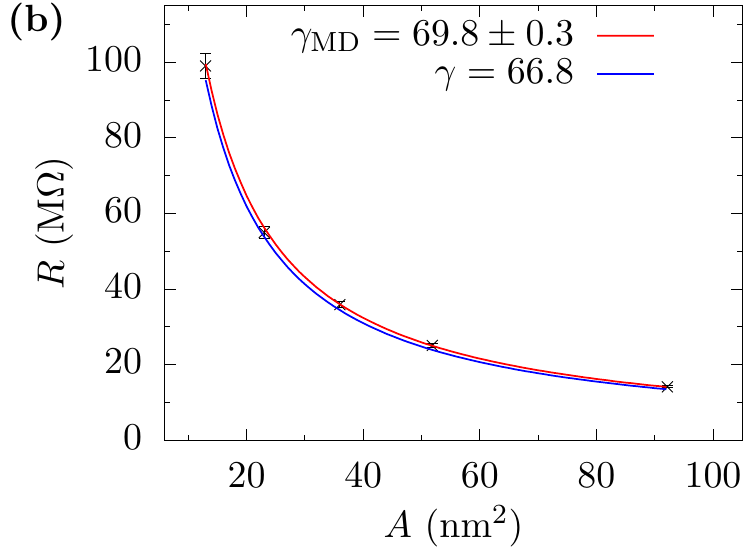}
\caption{\label{bulk}(a) Resistance versus simulation cell height and (b) resistance versus cross-sectional area for a bulk solution of 1 mol/L KCl. The red line gives $\gamma_\mathrm{MD}$ as the best fit to $R=\gamma_\mathrm{MD}h/ A$, and the blue line shows standard resistivity $\gamma=1/ne\left(\mu_\K+\mu_\Cl\right) \approx 67$ M$\Omega \cdot$nm. The error bars are $\pm 1$ BSE.}
\end{figure}

We calculate the bulk resistivity from our MD simulations using a cell without the graphene membrane/pore. The standard value of the bulk resistivity is $\gamma=1/ne(\mu_\K+\mu_\Cl) \approx 67$ M$\Omega \cdot$nm. The value from MD is $\gamma_\mathrm{MD}\approx 70$ M$\Omega \cdot$nm, as shown in Figure \ref{bulk}. It is to be noted that the actual value of resistivity of 1 M KCl at room temperature observed in experiments is $\gamma_{\mathrm{exp}}\approx 90$  M$\Omega \cdot$nm. At the high concentration of KCl (such as 1 mol/L), the conductance deviates from the linear expression, $\gamma=1/ne(\mu_\K+\mu_\Cl)$. However, the MD results give the conductance according to the linear expression.

\begin{figure}
\includegraphics{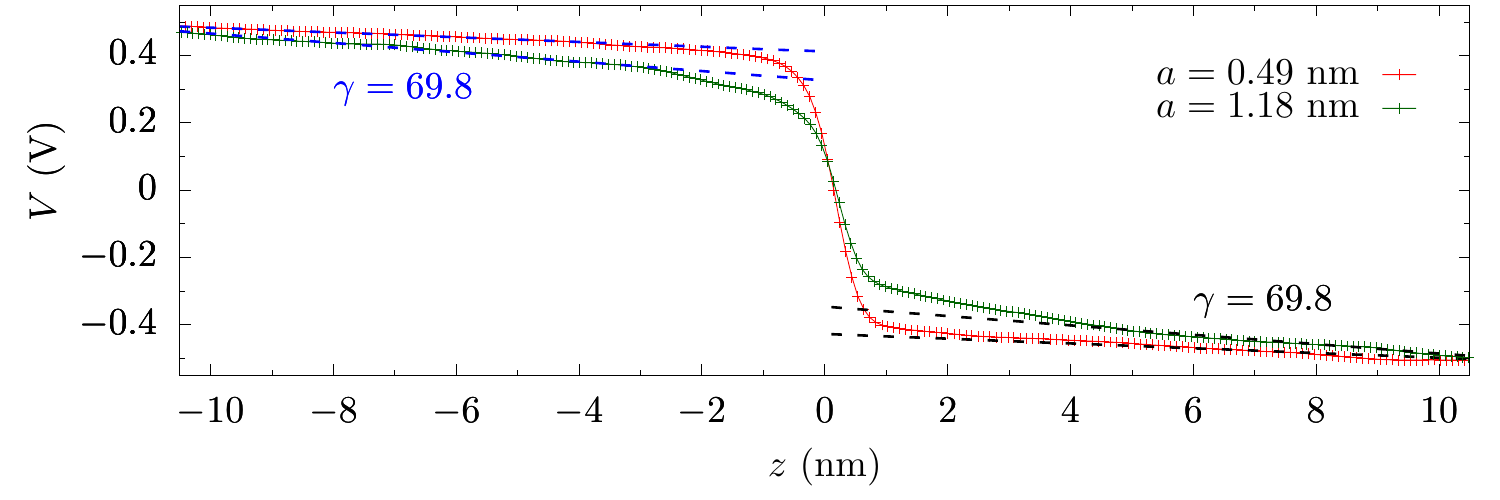}
\caption{\label{pot} Potential $V$ versus  the $z$-distance averaged over a cylindrical region of radius $\rho=1$ nm for $\rp=\ra$ nm (red line) and radius $\rho=2$ nm for $\rp=\rb$ nm (green line). The slope of potential drop is constant and equal to bulk resistiviy beyond $|z| \gtrsim L/2$ (with $L=9.6$ for $\rp=\rb$ nm and $L=7.2$ nm for $\rp=\ra$ nm). But within $|z| \lesssim L/2$ the slope increases initially very slowly and then very rapidly near the pore ($z=0$). The error in the potential is comparable to the size of data markers.}
\end{figure}

In Figure~\ref{pot} we plot the potential drop along the $z$-direction when 1 V potential is applied across the graphene membrane. At larger distances, $\Delta z$, from the pore, the potential drop, $\Delta V$, is proportional to the bulk resistivity $\gamma$, since

\begin{align}
\Delta V=I\Delta R= \frac{I \gamma\,\Delta z}{A},
\end{align}
where $\Delta R$ is the resistance of the region away from the pore.

\newpage

\bibliography{reference}